# Surface morphological evolutions on single crystal films by strong anisotropic drift-diffusion under the capillary and electromigration forces

Tarik Omer Ogurtani[a] and Aytac Celik[b]

Department of Metallurgical and Materials Engineering, Middle East Technical

University, 06531, Ankara, Turkey

The morphological evolution of voids at the unpassivated surfaces and the sidewalls of the single crystal metallic films are investigated via computer simulations by using the novel mathematical model developed by Ogurtani[1] relying on the fundamental postulates of irreversible thermodynamics. The effects of the drift-diffusion anisotropy on the development of the surface morphological scenarios are fully explored under the action of the electromigration (EM) and capillary forces (CF), utilizing numerous combinations of the surface textures and the directions of the applied electric field. The interconnect failure time due to the EM induced wedge shape internal voids and the incubation time of the oscillatory surface waves, under the severe instability regimes, are deduced by the novel renormalization procedures applied on the outputs of the computer simulation experiments.

---

[a] Corresponding author. Tel.:+90-312-210-2512; fax: +90-312-210-1267; e-mail:ogurtani@metu.edu.tr

url: http://www.csl.mete.metu.edu.tr

[b] e-mail: e104548@metu.edu.tr

## I. INTRODUCTION

So many years, the subject of capillary-driven morphological evolution at surfaces and interfaces of condensed matters, especially under the action of applied force fields such as electrostatic and thermo-mechanical stress systems, has represented a challenging theoretical problem in materials science, without having exposed to any robust non equilibrium thermodynamic or statistical mechanics treatments. This awkward situation is started to change very recently[2, 3] because of the sub-microscopic nature of the electronic devices that has pushed the surfaces and interfaces into the front lines as primary agents in the determination of the catastrophic failure of interconnect thin metallic lines used in the microelectronic industry.

Theoretical studies of interconnect surfaces under the electromigration force have also revealed a variety of morphological scenarios. Krug and Dobbs[4] and Schimschak and Krug[5] showed that a crystal surface could be destabilized by an external EM field in a material having anisotropic adatom surface diffusivity. Their linear instability analysis (LISA) strictly relies on the uniformly tilted surfaces and/or the small slope approximations, respectively. They also assumed that the field is almost constant along the surface. Nevertheless, it may be instructive in exploring the several aspects of the morphological instability of crystal surfaces and faceting transitions. In later studies, Schimschak and Krug[6], Gungor and Maroudas[7], and Ogurtani and Oren[8] put more emphasis on the crucial role of the surface diffusion anisotropy and the crystalline texture in the development of the awkward morphological variations on the preexistent edge or internal voids causing catastrophic electrical break down. In all these studies, the role of the boundary conditions on the dynamical behavior of the surfaces is under estimated. Gungor *et al.*[7] and Ogurtani and Oren[8, 9] used rigid sidewalls attached to the edge void, which made the interconnect sidewalls partially inactive. Schimschak and Krug[6] used active edge surface properly, but they continue utilizing periodic boundary conditions in addition to the constant voltage application, which also hinder the fine details of the process, especially during the estimation of the effect of the current crowding.

Only very recently, Ogurtani and Akyildiz[10, 11] have considered the profound effects of the reflecting and/or free-moving boundary conditions on the EM induced grain boundary grooving (GBG) and cathode voiding, in their computer simulation studies. These computer experiments showed irrevocably that in the electromigration dominating regime the GB voiding can be completely arrested by the applied current above the well defined threshold level.

Bradley[12] examined the effects of electromigration on the dynamics of corrugated interconnect-vapor interfaces using the multiple scale asymptotic analysis, by neglecting the capillary effects. This novel and very powerful technique was first introduced by Drazin and Johnson[13] to deduce the governing Korteweg-de Vries equation (KdV) equation for the irrotational 2-D motion of an incompressible and inviscid fluid. Bradley's[12] results are very interesting and predicting that the compressed and highly stretched solitary wave travels on the surface of a current carrying metallic thin film, having isotropic diffusivity, in the direction of the applied EM field. The propagation velocity and the width of the solitons decrease with increasing amplitude. These observations are in contrast to the behavior of ordinary solitary waves in shallow waters,[14, 15] which are also governed by a KdV equation. The most interesting work on the EM induced edge stability in single-crystal metal lines was carried by Mahadevan *et al.*[16] They employed a novel phase field (PF) technique to study this moving boundary problem numerically by assuming that the mobility of adatoms is anisotropic and having four-fold symmetry with $45^o$ tilt angle.

The most recent and extensive computer simulations are performed by Celik[17] on the finite size Gaussian shape edge hillocks and voids as initial data to induce Solitary waves without putting any physical and mathematical constrains on the model as advocated by Ogurtani[1] and Ogurtani and Oren[8]. These simulation experiments have irrevocably proved that even at the presence of strong diffusional anisotropy, the solitary waves (kinks or solitons and even a train of saw tooth waves) can be generated in the EM dominating regime, if one of the principal axes of the diffusivity dyadic has a special and irreducible orientation with respect to the applied electric field intensity vector.

*In situ* high voltage scanning electron microscope (HVSEM) examination of the accelerated EM tests performed by Marieb *et al.*,[18] clearly showed void nucleation at the sidewalls of the Al-1% Si lines at the metal-passivation interface. These voids grew into the line in an angled slit-like fashion, then begun to grow into a wedge-like shape. Several interesting phenomena were also observed that were not explained, such as, the liquid-like flow material during void coalescence and the dissolution of voids on the verge of breaching the line.

All these observations are motivated us to program, execute and analyze robust computer simulation experiments on the different combinations (192-variant) of the surface textures and tilt angles, the degree of anisotropy and EM wind intensity parameters by a novel mathematical model.

## II. PHYSICAL AND MATHEMATICAL MODELING

The evolution kinematics of surfaces or interfacial layers (simply or multiple connected domains) may be described by the following well-posed moving boundary value problem in 2-D space for the ordinary points, in terms of normalized and scaled parameters and variables.

$$\bar{V}_{ord} = \frac{\partial}{\partial \bar{\ell}}\left[D(\theta,\phi;m)\frac{\partial}{\partial \bar{\ell}}\left(\Delta\bar{g}_{vb} + \chi\bar{\vartheta} + \bar{\gamma}(\hat{\theta},\phi;m)\bar{\kappa}\right)\right] - \bar{M}_{vb}\left(\Delta\bar{g}_{vb} + \bar{\gamma}(\hat{\theta},\phi;m)\bar{\kappa}\right) \quad (1)$$

where $\bar{\kappa}$ is the local curvature and is taken to be positive for a convex void or a concave solid surface (troughs), $\bar{\gamma}(\hat{\theta},\phi;m) \equiv \{\gamma(\hat{\theta},\phi;m) + \partial^2\gamma(\hat{\theta},\phi;m)/\partial\hat{\theta}^2\}$ is the angular part of anisotropic surface specific Gibbs free energy. $\hat{\theta}$ and $\phi$ are, respectively, the angles between the line normal and the principal axis of the dyadic with respect to the applied electric field intensity vector in 2-D space. Similarly, the angular part of anisotropic diffusion may be represented by $D(\theta,\phi;m) = 1 + A\cos^2[m(\theta-\phi)]$, where $\theta$ is the angle between the tangent vector of the line profile and the electric field intensity vector, $A$ is the anisotropy constant, which may be a few orders of magnitude, $2m$ is the degree of fold of the symmetry (zone) axis and $\bar{\ell}$ is the curvilinear coordinate along the surface (arc length). $\Delta\bar{g}_{vb} = (\bar{g}_v - \bar{g}_b)$ denotes the Gibbs free energy of transformation ($\Delta\bar{g}_{vb} < 0$ evaporation or void growth), which is the Gibbs free energy difference between the realistic void phase (vapor) and the bulk matrix, and it is normalized with respect to the minimum value of the specific surface Gibbs free energy of the interfacial layer denoted by $g_\sigma^o$. $\chi$ is the electron wind intensity parameter, $\bar{\vartheta}$ is the normalized electrostatic potential generated at the surface layer due to the applied electric field intensity.

In above formula the surface drift-diffusion, which may be represented by an angular dependent post factor, $D(\theta,\phi;m)$ has been taken as anisotropic. On the contrary, for the time being the specific Gibbs free energy of the interfacial layer has been assumed to be isotropic. In this relationship, the bar sign over the letters indicates the following scaled and normalized quantities:

$$\bar{t} = t/\tau_o, \qquad \bar{\ell} = \ell/\ell_o, \qquad \bar{\kappa} = \kappa\,\ell_o, \qquad \bar{w}_o = w_o/\ell_o, \qquad \bar{L} = L/\ell_o \qquad (2)$$

$$\Delta\bar{g}_{vb} = \frac{\breve{g}_{vb}\ell_o}{g_\sigma^o}, \qquad \bar{\vartheta} = \vartheta/(E_o\ell_o), \qquad \chi = e|\hat{Z}|E_o\ell_o^2/(\Omega_\sigma g_\sigma^o) \qquad (3)$$

where, $E_o$ denotes the electric field intensity directed along the specimen longitudinal axis, $e|\hat{Z}|$ is the effective charge, which may be given in terms of the atomic fractions, $x^i$ by $\hat{Z} = \sum_i x^i \hat{Z}^i$ for multi-component alloys. In the present study the generalized mobility, $\hat{M}_{vb}$ associated with interfacial displacement reaction taking place during the surface growth process (adsorption or desorption) is assumed to be independent from the orientation of the interfacial layer in crystalline solids. It is normalized with respect to the minimum value of the mobility of the surface diffusion denoted by $\hat{M}_\sigma$. They are given, respectively by: $\hat{M}_\sigma = \left( \tilde{D}_\sigma h_\sigma / \Omega_\sigma kT \right)$ and $\bar{M}_{vb} = \left( \hat{M}_{vb} \ell_o^2 \right) / \hat{M}_\sigma$. Where, $\bar{\Omega}_\sigma$ is the mean atomic volume of chemical species in the void surface layer. $\tilde{D}_\sigma$ is the isotropic part (i.e., the minimum value) of the surface diffusion coefficient. In the formulation of the problem, we have adapted the convention such that the positive direction of the motion is always towards the bulk material whether one deals with inner voids or outer surfaces or interfaces.

In above expressions, $\hat{M}_{vb}$ may also be identified as a reaction rate constant associated with the phase transformation denoted symbolically by $v \Leftrightarrow b$. In general, this phenomenological mobility is not only strongly dependent on the temperature, but also on the orientation of the surface layer and the applied stress system.[19] It may be formulated according to the activated complex theory of chemical reaction rates.[20]

In the earlier description,[3] we have tried to scale the time and space variables $\{t, \ell\}$ in the following fashion; first of all, $\hat{M}_\sigma$, an atomic mobility associated with mass flow at the surface layer, is defined previously, and then a new time scale is introduced by $\tau_o = \ell_o^4 / \left( \Omega_\sigma^2 \hat{M}_\sigma g_\sigma \right)$, where $\ell_o$ is the arbitrary length scale, which is for the present simulation studies chosen as $\ell_o = 2w_o/3$, where $w_o$ is the half width of the interconnect specimen.

## III. THE RESULTS AND DISCUSSION

In our present computer simulation studies, it is assumed that the sample was sandwiched with the top and bottom high resistance (TiAl$_3$, TiN, et cetera) coatings, which together with the substrate constitute diffusion barrier layers. It is also assumed here that only the one edge (sidewall) of the interconnect line is subjected to the surface drift-diffusion, and it is exposed to environment whose conductivity is neglected in this study. In certain cases,[21] in which the upper surface of the unpassivated interconnect becomes predominant path for the drift-diffusion; the results of our computer simulation may still be applied by modifying the line width parameter, $2w_o$ with the line thickness denoted by $h_o$.

All together 192 different combinations of the surface textures, drift-diffusion anisotropy coefficients, and electron wind intensity parameters are subjected to this simulation work on the edge voids and the hillocks. In all these cases, the direction of the electron wind is chosen along the longitudinal axis of the interconnect line to sweep the unperturbed sidewalls and upper and lower surfaces of the test piece. We have employed mainly four different electron wind intensity parameters, ($\chi = 5, 10, 25$ and $50$) in the electromigration dominating regime (EMDR), which covers from the moderate up to the high current densities ($j = 10^9 - 10^{12}$ A/m$^2$) that are mostly utilized in normal and accelerated laboratory test studies. As an initial data for the surface topology, a Gaussian shape void is introduced on top of the otherwise perfectly smooth flat edge surface. We have also adapted large aspect ratios, $\beta = 20-30$ with reference to the halve-width of the

void or hillock present at the surface, and utilized quasi-infinite boundary conditions such as that the surface curvature and their higher derivates are all equal to zero at the anode and cathode edges. The constant and uniform electric field is applied to the specimen as a particular solution (initial data), which insures that there is steady flow of atomic species from the cathode end to the anode edge of the specimen. In order to sustain a constant current condition all along the simulation experiment, the induced electric field (the complementary solution) due to surface disturbances and the EM induced internal voids are calculated by modified indirect boundary element method (MIBEM), using special Neumann boundary conditions which leaves the cathode and anode edges completely open for the constant current flow, and seals off the sidewall (edges) surfaces, and the EM induced internal voids from any current penetration.

In order to explain the some general trends in the morphological evolution of the disturbances on the otherwise flat solid surfaces, references are made to the linear instability analysis (LISA) recently developed by Brush and Oren.

**Surface void configurations:**

The morphology of an initially perfectly flat surface having a perturbation in the shape of Gaussian edge-void is demonstrated in Fig. (1), where the positive direction of electric field is from the left (anode) to the right (cathode). The scaled interconnect width is denoted as $\bar{w}$ and the void depth and the specimen length are given by $\bar{a}$ and $\bar{L}$, respectively. These are all scaled with respect to the arbitrary length denoted by $\ell_o$. Three different crystal planes, $\{110\}$, $\{100\}$ and $\{111\}$ for the surface of the single crystal thin films having a fcc crystal structure, are considered, which may be characterized by two, four and six fold symmetry zone axes, respectively.

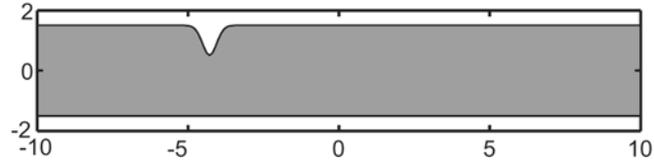

**FIG. 1:** Front view of a thin metallic single crystal film with a preexistent edge void just before the onset of the application of the electric field.

The macro behavior of the edge-void evolution kinetics in terms of the applied electron wind intensity and the degree of diffusion anisotropy is analyzed for each texture configuration in the following sub-sections, and the results are summarized in Tables (I, II and III), for the two-fold, four-fold and six-fold symmetry planes. In general, the tilt angle between the direction of the electric field and the one of the principal axis of the diffusion dyadic is defined in the semi-closed interval.

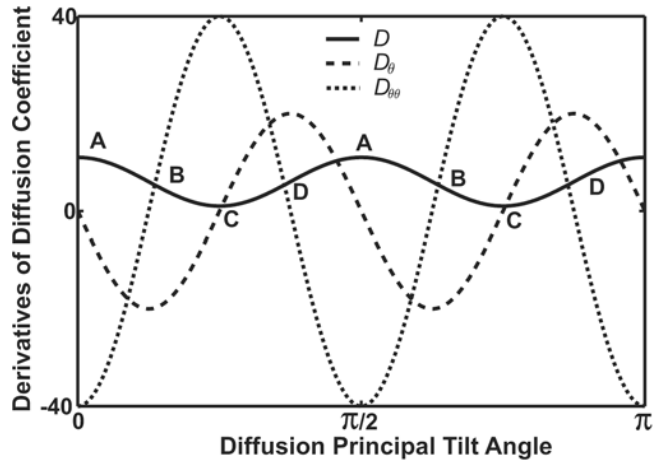

**FIG. 2.** Anisotropic diffusion constant and its higher order derivatives for the four-fold symmetry plane $\{100\}$. A: very unstable (the wedge shape fatal internal void); B: very stable; C: for low EW-decay; for high EW-solitary wave train; D: very unstable (oscillatory waves).

In Fig. (2), the anisotropic diffusion coefficient and its first two higher derivatives are plotted against the tilt angle, for the four fold-symmetry planes of a fcc structure, which will be used as a prototype later systematically in our discussions on the instability analysis of the finite surface perturbations. Since the present computer simulation experiments have been devoted for the electromigration dominating regime, namely $\chi > 1$, one can easily differentiate two distinct domains, (dissipative and regenerative) in terms of the tilt angle only without going any complications with the capillary effects. There are two important exceptions, which are related to the transition stages taking place while passing from the stability to the instability region or vice versa, and they cannot be predicted by the LISA theory. These critical transitions in the tilt angles are corresponding to the extremum in the anisotropic diffusion coefficient denoted by $\phi = 0$, and $\phi = \pi/m$, and plus their periodic extensions. According to the results of our computer simulations performed in the electromigration dominating regime, the stability and instability regimes for the finite amplitude perturbations may be defined by the following open intervals for the tilt angles: ($0 < \phi < \pi/2m$) and ($\pi/2m < \phi < \pi/m$), and plus their periodic extensions represented by $\{n\pi/m\}$, where $n \leq m$ is a set of positive integer numbers, respectively.

*i. Two Fold Crystal Symmetry, $\{110\}$ Planes in fcc:*

The results of the extensive computer experiments performed on the sidewalls of the $\{110\}$ single crystal thin film surface, using various EW intensity parameters, tilt angles and the diffusion coefficient anisotropy constants (*A=5* and *A=10*) are outlined in Table

I, in a systematic fashion. The interconnect having two fold symmetry surface with a zero degree tilt angle $\phi = 0^o$ with respect to the electric field direction tends to transform the Gaussian shape edge-void into the slit like shape, stretched to the windward direction with about $45^o$ inclination as also observed by Mahadevan *et al.*[16] Eventually, the tip of this slit type void breaks down and generates an EM induced internal void having various different in forms. The close inspection of the anisotropic diffusivity expression presented earlier shows that for this tilt angle, the diffusion rate is maximum along the applied electric field, $D_\phi(0, \phi \to 0, m) = 0$ and $D_{\phi\phi}(0, \phi \to 0, m) < 0,$ where $m = 1$. According to the first order linear instability analysis (LISA) of the Ogurtani[2] governing equation by Brush and Oren for this tilt angle one should be in the stability regime since the contribution from the electromigration to the growth rate constant denoted by $\Gamma$ becomes equal to zero, and the remaining term due to the capillary force is negative (dissipation), and only contributes to the stability. Therefore, at the transition points between the stability and the instability regions such as the present case, one should be very careful in using LISA theory. The system in this configuration is extremely unstable, and the newly created internal EM induced void is strongly affected by the intensity of the electron wind.

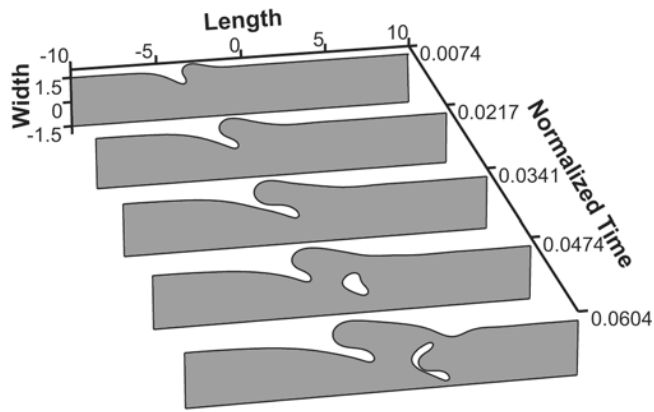

**FIG. 3.** The morphology evolution for an edge void in the electromigration dominating regime. Deformation into a slit like shape stretched towards the windward direction about $45^o$ inclination is followed by the detachment of an internal void while all together moving towards the cathode side. $V_S = 140$, and $V_v = 136$,.

At low electron wind intensities $\chi \leq 10$, as can be seen from Fig. (3), at first a large size slit shape internal void forms, which slowly changing its shape into hook form while migrating towards the lower surface firmly attached to the substrate. Finally it hits the lower interface of the sample creating fatal and deep crack. This process doesn't stop there, and rather repeats itself till the complete electrical breakdown takes place. Where, the surface disturbance and the EM induced void drift velocities are denoted by $V_S$ and $V_v$ ,respectively, and are given in terms of normalized units.

On the other hand, at the moderate and high electron wind intensities $\chi = 25 - 50$ somehow partial stability appears, the edge void after transforming its shape, the almost cylindrical and small in size internal void forms, which slowly drifts towards the cathode end of the sample as illustrated in Fig. (4). The size of these internal voids seems to be inversely proportional with the electron wind intensities. The incubation times for the creation of EM induced internal voids strongly depend on the applied electric field. At high electric fields, such as $\chi = 50$, the small size void forms with relatively short incubation time. For the moderate electron wind intensities, the incubation time for the void creation becomes longer.

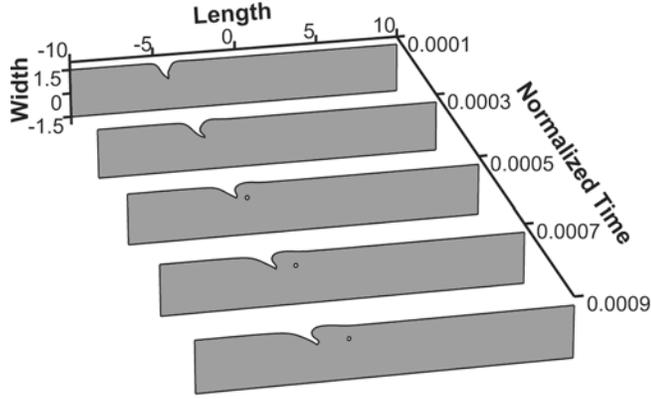

**FIG. 4.** At very high electron wind intensities, the edge void shows minor change in shape while generating an internal round void having small in size, but moving with a great speed towards the cathode end. $V_S = 1565$, and $V_v = 3525$.

In Fig. (5), the case of $\phi = 45^o$ degrees tilt angle is demonstrated, where the edge-void that is initially established on the flat surface starts to disappear after turning on the applied electric field. The intensity of the applied electric field and the degree of diffusion anisotropy affect only the total decay time of the edge-void. At low and moderate electron wind intensities $\chi \leq 25$, the edge void disappears without leaving any trace behind. This behavior in EM dominating regime characterizes the maximum dissipation rate, which corresponds to the inflexion point on the ($D(0,\phi,m)$ vs $\phi$) plot, that implies following mathematical connection: $D_\phi(0,\phi,m) < 0$ and $D_{\phi\phi}(0,\phi,m) = 0$.

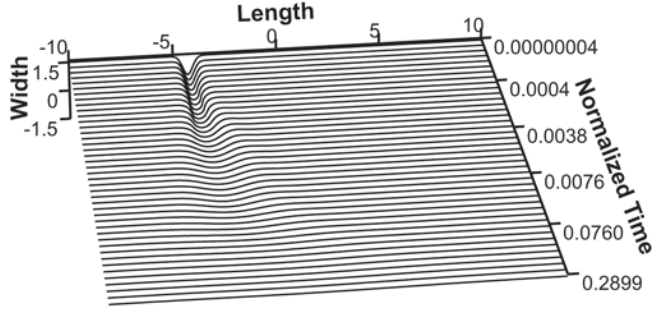

**FIG. 5.** The surface perturbation due to the edge void under the small EW intensity is completely disappearing without leaving any trace. $V_s$: 18.7; The decay time for 95% reduction: 0.023, in normalized units.

At very high electron wind intensities, $\chi \geq 50$, interconnects having 45 degrees tilt orientation show EM induced voiding having extremely short incubation time, which is about $\bar{t}_v \cong 0.0005$. The surface velocity is $V_s \cong 1261$, and the void drift velocity is about $V_v \cong 1939$ in normalized units, respectively. In this extreme EM dominating regime, the edge-void perturbation is completely disappearing from the surface after popping up a few, small and short leaving EM voids.

When the tilt angle changes to 90 degrees, the edge-void changes its morphological evolution behavior completely, especially at low and moderate electron wind intensities. This is a typical and very unique regime as can be seen in Fig. (6), which can be characterized by $D_\phi(0,\phi) = 0$ and $D_{\phi\phi}(0,\phi) > 0$, in the anisotropic part of diffusion dyadic plotted in Fig. (2). This configuration, in which the diffusivity is still minimum in the direction of the electron wind that is sweeping the sidewall surface, corresponds to the transition stage between dissipation and regenerative regimes labeled as 'C' in that figure. Here one observes the formation of a Solitary Wave by the transformation of the Gaussian shape into the wedge shape, having sharp drop on the leeside.

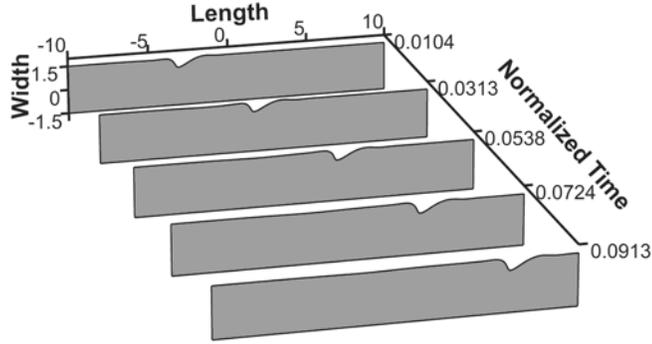

**FIG. 6.** The formation of a saw tooth shape solitary wave from the Gaussian shape edge void while migrating towards the cathode end, $V_S : 116$.

According to the data obtained by the computer simulations, the solitary wave position with respect to the elapsed time is given by the relationship $\bar{x} \cong 0.78(1+A)\chi\bar{t} + 0.02$, which is valid in the range of $\chi \leq 25$, where no internal void ejection takes place. This expression may be renormalized to obtain the following relationship:

$$x \cong 2.3(1+A)\frac{\tilde{D}_\sigma h_\sigma e |Z_\sigma|}{kTw}\rho_b J_b t + 0.007w, \quad \text{(Solitary Wave Displacement, } \chi \leq 25 \text{)} \quad (4)$$

This explicit expression shows that the solitary wave velocity is a function of the line width and the anisotropy coefficient. Since we don't have enough data for the amplitude and the line width dependence of the pre-constant in above expression, all we can say that it is inversely proportional with the height of the original edge void situated at the sidewall. This dependence is just opposite to those Solitary waves obtained from the solutions of the KdV or sine-Gordon equations. However, it is in complete agreement with the findings of Bradley,[12] as we mentioned in the introduction.

This Solitary wave doesn't have any conventional shapes deduced from the KdV equation or from the sine-Gordon equation). This Solitary wave, which is first time detected by us in computer simulation experiments has a saw tooth shape, and migrates towards the cathode edge with constant velocity without showing any indication of dispersion and dissipation of energy.

At very high electron wind intensity regime $\chi \geq 50$ combined with moderate anisotropy constant such as $A = 5$, the morphological evolution of the edge-void becomes extremely interesting, namely: the edge void changes its shape into the saw tooth form similar to the case that mentioned above. But now! it starts to emanate very small and rounded shape void from its lee side.

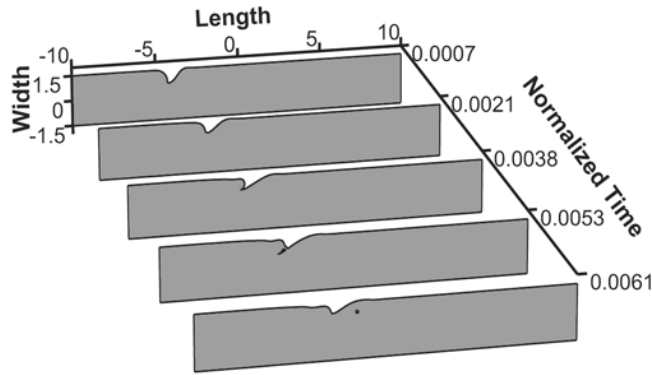

**FIG. 7.** The formation of a saw tooth solitary wave with the ejection of a round internal void at very high EW intensity. $V_S = 232$ and $V_v = 2199$.

This EM induced tiny void, as seen clearly in Fig. (7), starts to drift towards windward side, and passes it's slowly dragging mother, and then continues in traveling in the direction of the cathode edge while tracing rather straight path. The incubation time for EM induced void formation is a decreasing function of the diffusion anisotropy coefficient denoted by *A* in Ogurtani governing formula. The solitary surface wave drift velocity is given exactly by Eq. (7). The EM induced void velocity is also obtained, which is about factor of 9.5 faster then the surface solitary wave drift velocity. The behavior of the solitary wave at high anisotropy constant $A \geq 10$ is quiet different. The ejection of the first internal void starts much earlier than the previous case, and in addition the system pops out more internal voids like a continuous source. The surface drift velocity shows some erratic behavior, which cannot be explained by the above given analytical expression anymore.

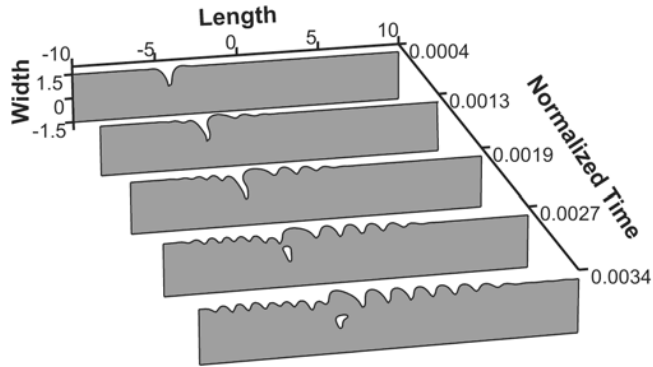

**FIG. 8.** Morphological evolution of the surface edge at the critical tilt angle showing extreme instability by producing oscillatory waves in concurrently ejecting an internal void due to the extremely high EW intensity. $V_v = 880$.

For the development of the surface topography, $\phi = 135$ degrees orientation of the principal axis of the diffusivity dyadic with respect to the direction of the electric field is very critical. The strong instability associated with this tilt angle is characterized by the inflexion point labeled as 'D' in the diffusivity plot in Fig. (2), which is specified by $D_{\phi\phi}(0,\phi,m) = 0$ and $D_{\phi\phi\phi}(0,\phi,m) < 0$. The preset edge void perturbation on the surface becomes unstable immediately after turning on the electric field. At first the Gaussian edge-void transform into a wedge shape having sharp edge on the windward side as may be seen in Fig. (8), and later the formation of oscillatory wave on the lee and windward sides appears, which spreads in all directions in space with deceasing in amplitude. The wedge shape edge-void finally ejects a small but rounded internal void to the bulk region, and it disappears completely at the background oscillatory waves. This phenomenon occurs regardless the intensity of the electron wind as long as $\chi \geq 1$, below which it may be stabilized because of the predominating effect of the capillary forces over the electron winds.

### ii. Four Fold Crystal Symmetry, $\{100\}$ planes in fcc:

In Fig. (9), the effect of the low electron wind intensity $\chi \leq 10$ on the surface topology of an interconnect line with fcc structure having four fold crystal symmetry, and oriented with a tilt angle of $\phi = 0$ degree is illustrated. The originally Gaussian in shape edge-void on the surface starts to evolve into a kink shape disturbance, which shows continues growth in size while drifting towards the cathode end of the interconnect line.

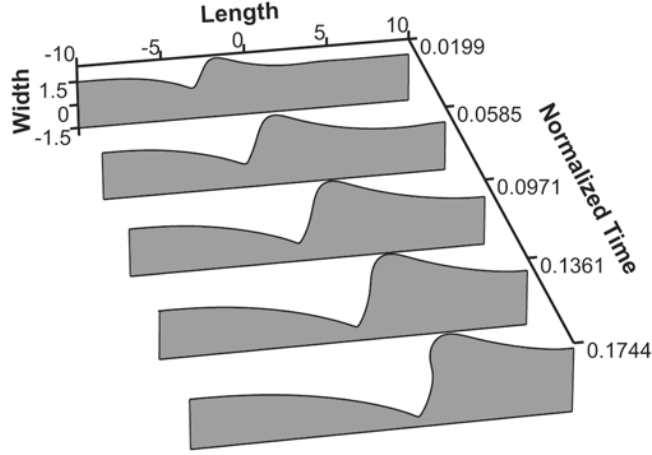

**FIG. 9.** Morphological evolution of the sidewall void to a kink shape disturbance, which changes dramatically its shape into slat just before reaching the lower side of the interconnect line. At the final stage, the fatal failure of interconnect takes. $V_S = 36$.

A close inspection of the anisotropic diffusivity plot presented in Fig. (2) shows that the diffusion rates are maximum not only in the direction of electric wind sweeping the sidewall surface layer, but also in the directions perpendicular to the sidewall surface, which are also called as the periodic extensions denoted by letter 'A'. This secondary feature just mentioned previously is the prime factor for the extreme instability of the surface perturbations. One has exactly same condition on the diffusivity dyadic as mentioned above for the set of {110} planes, namely; ($D_\phi(0, \phi \to 0, m = 2) = 0$ and ($D_{\phi\phi}(0, \phi \to 0, m = 2)$ <0). This point also corresponds to the transition state between the stability to the instability regimes, and in addition having maximum diffusivity in the direction of perpendicular to the sidewall surface.

As may be seen clearly in Fig. (10), at the moderate electron wind intensities $\chi \approx 25$, the kink shape surface disturbance breaks down and creates very large wedge shape internal void, while becoming very close to the opposite sidewall of the interconnect.

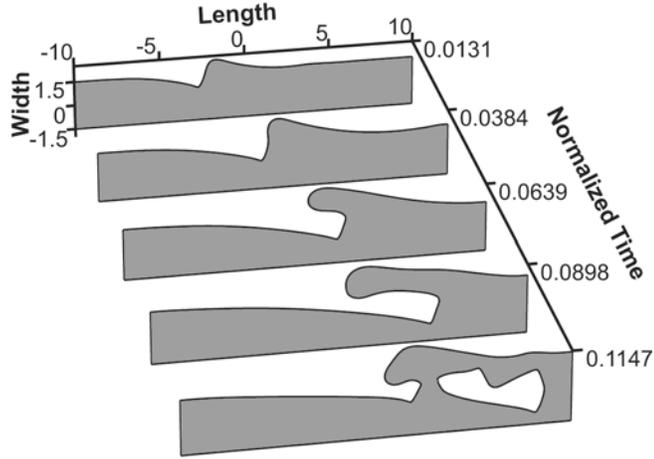

**FIG 10.** Morphological evolution of an edge void on the sidewall of the (100) plane, during the detachment of a large piece of internal void. At the final stage this causes the fatal failure of the interconnect line. $V_S = 98$ and $V_v = 151$.

On the other hand, at extremely high electron wind intensities $\chi \geq 50$, the lower edge of the kinked shape surface void grows so much that eventually hits the opposite sidewall before having enough time to break down into internal daughter voids. Therefore this type textures are very critical and cause an open circuit failure by reaching the other edge. The failure time depends on the degree of anisotropy in diffusion coefficient and the electron wind intensity.

In Fig. (11), the failure times of single crystal thin film interconnects exposed to the electron wind at zero tilt angle are plotted with respect to the EW intensity parameters on a double logarithmic scale.

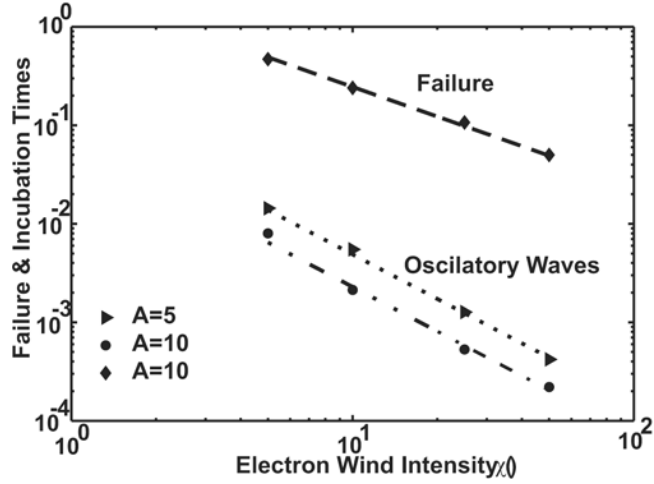

**FIG. 11.** The failure and oscillatory wave incubation times for the zero and the 60° tilt angles associated with the 4-fold symmetry plane {100} are plotted as a function of the EW intensity on the double logarithmic scale, respectively. The anisotropy constant *for* the failure time is A=10. For the incubation time, we have two different sets of data, namely; A=5 and A=10.

As we discussed previously all these samples have {100} surface orientation, and $\phi = 0$ tilt angle show the fatal failures caused by the wedge type internal void as demonstrated in Fig. (10). According to the linear regression analysis one finds the following relationship: $\bar{t}_F \cong 27\left[(1+A)\chi\right]^{-1}$ in terms of scaled and normalized quantities. The renormalization of this expression yields the following very interesting formula:

$$t_F \cong 27 \frac{\ell_o^2}{(1+A)} \left[\frac{\tilde{D}_\sigma^o h_\sigma}{kT} e|Z|E_o\right]^{-1} \cong 3 \frac{w^2}{(1+A)} \left[\frac{\tilde{D}_\sigma^o h_\sigma}{kT} e|Z|\rho_b J_b\right]^{-1}, \text{(sidewall failure)} \quad (5)$$

where, $w$ is the full width of the metallic film, $\rho_b$ is the bulk specific resistivity, and $J_b$ is the applied current density, which is kept constant during the computer simulation experiments. This expression shows that if the line failure occurs due to the EM induced voiding by the sidewall surface diffusion than the failure time depends upon the line width quadratically having a current exponent $n = -1$. The relationship given in Eq. (5) has exactly same functional dependence on the system parameters and the external variables compared to the expression for the failure time associated with the cathode voiding by the sidewall surface diffusion as very recently deduced by Ogurtani and Akyildiz[10] from the analysis of their computer simulation experiments (i.e., Eq. (21) in Ref.10). Both expressions have identical functional form in the case of isotropic surface diffusion $A = 1$, but they have quite different pre-constants, namely; instead of 3 in Eq. (5), the cathode failure time has 0.06, which is almost two orders of magnitude smaller. That is the main reason behind the observations of the cathode failure mechanisms in polycrystalline interconnect lines in majority cases rather than the interconnect sidewall failure by the EM induced internal voids in practice, in the case of isotropic or moderate anisotropic interfacial diffusivities, $A \leq 50$.

We also investigated the case where the tilt angle is $\phi = 30$ degrees, and the general evolution behavior is demonstrated in Fig. (12) for low and moderate electron wind intensities $\chi = 5-10$.

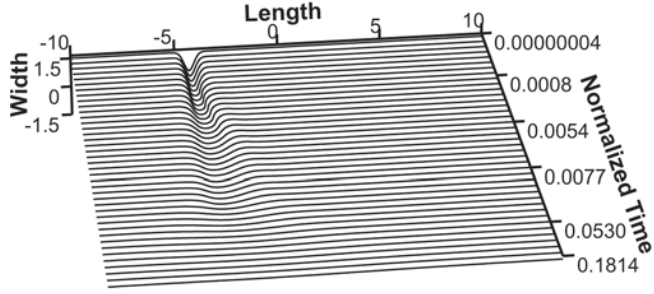

**Fig. 12.** Complete disappearance of the preexisted edge void on the sidewall of a (100) surface after the application of electrostatic field.

The void shows some shape variations while decaying, and finally disappears completely without leaving any trace on the otherwise flat surface. However, at very high electron winds $\chi \geq 50$ according to Fig. (13) a round shape daughter void may be created, which migrates towards the cathode edge. There may be even multiple daughter voids emanating from the same source at later times. This behavior in EM dominating regime characterizes the maximum dissipation rate, which corresponds to the inflexion point on the ($D(0,\phi,m=2)$ vs $\phi$) plot denoted by letter 'B' in Fig. (2), which implies following general mathematical connection; $D_\phi(0,\phi,m=2) < 0$ and $D_{\phi\phi}(0,\phi,m=2) = 0$, where the exact tilt angle is given by $\phi \equiv \pi/8 = 22.5°$.

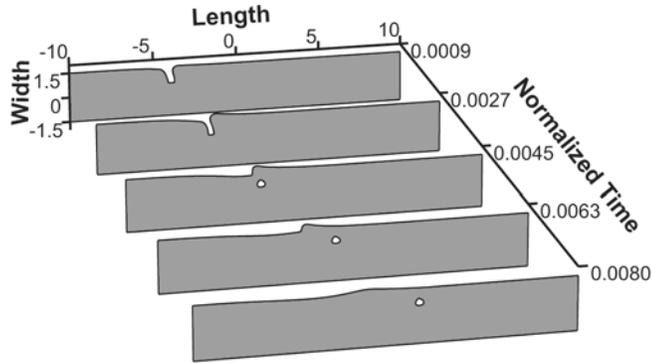

**FIG. 13.** The morphological evolution of an edge void on the sidewall of a (100) plane under severe electromigration forces. Even the system under the absolute stability regime, still it rejects a round fast moving internal void, which moves towards the cathode end. $V_S = 341$ and $V_v = 1165$.

The effects of 45 degrees tilt on the edge surface topology may be studied in two different categories; the capillary effect dominating low electron wind intensity $\chi \leq 10$ regime, and the EM dominating moderate and the high electron wind intensity regime $\chi \geq 25$. At low current density regime, the edge void disappears completely without leaving any trace on its back. On the other hand the electromigration prevailing regime shows an unusual characteristics as can be seen in Fig. (14), which still can be characterized by $D_\phi(0,\phi,m=2)=0$ and $D_{\phi\phi}(0,\phi,m=2)>0$, in the anisotropic part of diffusion dyadic.

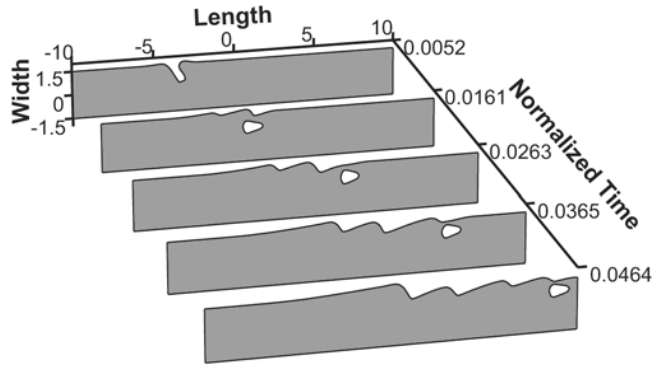

**FIG. 14.** Morphological evolution of (100) plane under the transition state exposed to the moderate electron wind intensity, which shows train of solitary waves stretched towards the windward side, and an ejected internal void moving towards the cathode end. $V_S = 177$, and $V_v = 326$.

However, one cannot observe any single soliton formation similar to the 2-fold case as it was studied earlier. In the present case, the edge-void changes its Gaussian shape by transforming into the slit-like configuration, and then emanates an internal void, which has almost perfect wedge form, pointing towards the windward direction. This wedge shape inner void migrates along the interconnect line towards the cathode end following a straight path. The mother edge-void after releasing the daughter void slowly transforms into saw tooth type surface undulations. This highly localized the package of train of few saw tooth undulations starts to spread by adding more waves by keeping the original wave form invariant but increasing the number. The whole wave package drags towards the windward direction while keeping close track with motion of the drifting daughter void.

When the tilt angle is changed to 60 degrees, the edge-void starts to decease by generating surface undulations in all directions as long as the electron wind intensity stays in the range of $\chi \leq 25$. The amplitudes of these waves become more pronounce and their waveform becomes saw tooth-like shape having relative steep windward edge. It seems that there is steady translational motion of this wave package towards the cathode end with constant growth in amplitudes. This maximum instability is characterized by the second inflexion point denoted by letter 'D' in the diffusivity plot presented in Fig (2), and it is defined as $D_{\phi\phi}(0,\phi,m) = 0$ and $D_{\phi\phi\phi}(0,\phi,m) < 0$, where $\phi = 3\pi/8 = 67.5^o$.

In Fig. (11), the incubation time of highly unstable oscillatory waves generated by the edge void on sidewall surfaces exposed to the electron wind at the $60^o$ tilt angle is plotted for two different diffusion anisotropy constants, namely; $A = 5$ and $A = 10$. The linear regression analysis of the data shows that the expression denoted by $\bar{t}_{OS} \cong 0.85(1+A)^{-1} \cdot \chi^{-1.5}$ gives very good fit. After the application of the renormalization procedure, one obtains the following expression for the incubation time for the oscillatory wave formation;

$$t_{OS} \cong 0.28w \left[ \frac{(1+A)\tilde{D}_\sigma h_\sigma}{kT\sqrt{\Omega_\sigma g_\sigma}} \right]^{-1} \left(e|Z|\rho_b J_b\right)^{-3/2}, \text{(oscillatory wave incubation)} \qquad (6)$$

This relationship shows that the current exponent for the incubation time for the oscillatory waves generation $n = -1.5$, and it depends on the line width linearly and on the anisotropy constant inversely. Other important peculiarity of this regime is its square root dependence on the specific surface Gibbs free energy, which means that the high capillary forces may have the healing effects by suppressing or delaying the formation of oscillatory instability waves, which covers the whole surface area creating extremely rough topology.

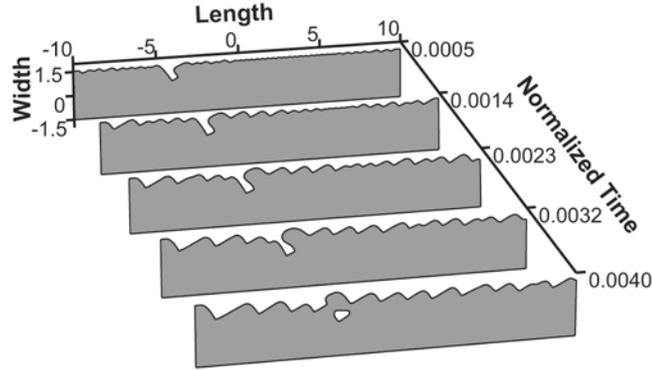

**Fig. 15.** Formation of oscillatory wave train accompanied by the ejection of EM induced internal void. $V_v \cong 1348..$

The situation is completely different for very high EW intensities $\chi \geq 50$, where one observes two distinct morphological variance, namely; low anisotropy constant region $A \leq 5$, and high anisotropy region, $A \geq 10$. For the low anisotropy region; the edge void first intrudes into the interconnect, which is followed by the ejection of an internal small round void, and than transforms into kink shape solitary surface wave drifting steadily towards the cathode end without changing its form. The solitary drift velocity is 307, and the induced void velocity is 2665, which is about a factor of time greater, in normalized units. In the case of high anisotropy, which is illustrated in Fig. (15) during the intrusion of the edge void inside the interconnect line, the oscillatory surface waves starts to appear and spreads in all directions, and finally the ejection of a internal wedge void takes place. These voids drift steadily towards the cathode edge, with a uniform velocity given by 1348 compared to the mother void speed of 345 in normalized units.

### iii. Six-Fold Crystal Symmetry, $\{111\}$ Planes in fcc:

The $\{111\}$ plane has the highest symmetry, 2m=6 compared to all other crystal planes in fcc structure. The orientation of the surface with zero tilt angles, $\phi = 0$ with respect to the electron wind direction at low current densities doesn't create very much trouble, because of the stabilization effect of the capillary forces. However, at moderate and very high EW intensities situation becomes very critical as may be seen in Fig. (15). Here one has a transition state from the instability to the stability regimes: $D_{\phi}(0,\phi,m) = 0$ and $D_{\phi\phi}(0,\phi,m) < 0$. The edge-void while dragging towards the cathode end, with an increasing intensity, changes its form into a kink-shape, which is somewhat tilted towards the windward side.

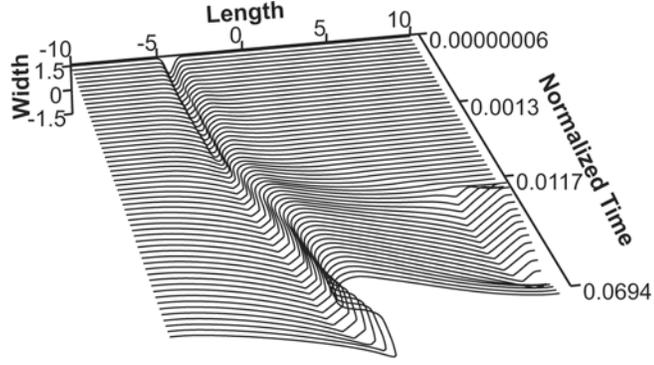

**FIG. 16.** Morphological evolution of an edge void at the sidewall of a (111) plane under the moderate EW intensity. The edge void changes its form becoming a wedge intrusion penetrating deep into the interconnect line finally reaching the opposite sidewall and causing the complete electrical breakdown . $V_S = 68$.

This form may cause an open circuit failure by reaching the other edge, but the failure time becomes very long compared to those interconnects, which have different surface textures such as {100} and {110} planes. The failure time of the interconnect is analyzed and it is found to be $\bar{t}_F \cong 19\left[(1+A)\chi\right]^{-1}$ in the range of $\chi \leq 25$ that it may be transform into the following expression by the renormalization procedure;

$$t_F \cong 2.1 \frac{w^2}{(1+A)} \left[ \frac{\tilde{D}_\sigma h_\sigma}{kT} e |Z_\sigma| \rho_b J_b \right]^{-1}, \text{ (sidewall failure time)} \qquad (10)$$

This relationship is almost identical to the expression Eq. (5) presented previously for the sidewall failure time of an interconnect line, having {100} surface plane and the zero tilt angle orientation with respect to the EW, by the EM induced wedge shape internal void. This peculiar situation indicates that the failure times associated with the morphological evolutions of the sidewall edge voids do not depend on the fine or the secondary features of the mechanism of the break down, whether it occurs by the direct hit coming from the extruded nose of the mother edge void or by the coalesce of the approaching daughter void towards the opposite sidewall.

At high electron wind intensities, $\chi \geq 50$, the situation somehow differs from above mentioned low and moderate EW intensity cases. Now, the edge void topological evolution shows two different and distinct trends for the low anisotropy constant $A \leq 5$ and the high anisotropy constant $A \geq 10$. In the case of the low anisotropy a small internal void forms shortly after turning on the electric field, and then this EM induced void drifts straight towards the cathode edge with very velocity $\bar{V} \cong 1554$. On the other hand for the large anisotropy, after the transformation of the original edge void into a kink-shape than it starts to extent deep into the interior and takes a bottleneck form. This bottleneck region finally breaks down due to the current crowding, and creates a very large internal void in wedge shape, which touches the opposite edge and causes fatal circuit shut down. Actually, a close inspection shows that in both regimes, the interconnect failure time obeys the same expression given by Eq. (5) or Eq. (7).

For 15 degrees tilt angle, two distinct behaviors are observed, namely low anisotropy constant $A \leq 5$ regime and the high anisotropy regime. In the anisotropy regime the edge void completely decays of if the electron wind intensity is in the range of $\chi \leq 5$, otherwise the edge-void regardless the intensity of the electron wind transforms into an inclined kink-shape and migrates towards the cathode end with constant amplitude having uniform velocity. For the high anisotropy case, the wave front becomes more and steeper on the leeward side, during drift as can be seen in Fig. (17). At the end, it may transform into a step like surface morphology. Where one has the following relationship: $D_\phi(0,\phi,m) < 0$ and $D_{\phi\phi}(0,\phi,m) = 0$, which corresponds to the maximum stability in electromigration dominating regime labeled as 'B' in Fig. (2).

The situation may be more critical at moderate and very high current densities $\chi \geq 25$, and low aspect ratios, because of this steady increase in intensity combining with steeping of the lee side, may cause substantial decrease in the cross section of the interconnect line and eventually circuit break down takes place by local joule heating. This behavior in EM dominating regime is not in accord with the prediction of the linear stability analysis, which claims that one should have maximum dissipation in this tilt angle for the (111) symmetry plane.

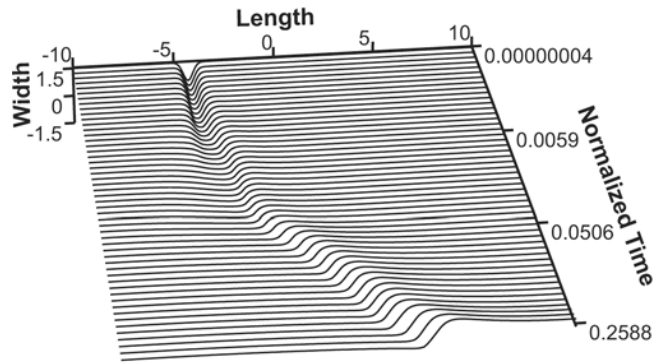

**FIG. 17.** The sidewall surface morphology evolution at low EWI clearly shows the kink shape solitary wave propagation towards the cathode end. The drift velocity of the solitary wave, $V_S$: 30 in normalized units.

Fig. (18) shows that at the 30 degrees tilt angle, the edge-void transforms into a kink-shape step at the surface for the low electron wind intensities -having a steep edge on the windward side- which may multiply on the windward side while the wave package train drifting towards the cathode end. It can be characterized by $D_\phi(0,\phi,m=2)=0$ and $D_{\phi\phi}(0,\phi,m=2) >0$, and labeled as 'C' in Fig. (2), which actually represent a transition stage between the dissipation to the regenerative regime, it should show some short of the solitary wave characteristics.

At moderate wind intensities combined with high anisotropy, a hip or hillock forms in front of the traveling edge-void, which grows steadily and eventually over hangs and traps the edge-void as such that the edge void becomes an interior void. This very unusual process is more pronounce and fast when the diffusion anisotropy coefficient is low, about $A \leq 5$ but the electron wind intensity is very high, $\chi \geq 50$. This self-trapped void becomes wedge shape and drags towards the cathode end. During the travel period, it may or may not touch the opposite edge of the interconnect depending upon the aspect ratio and the size of the initial Gaussian edge void with respect to the line width.

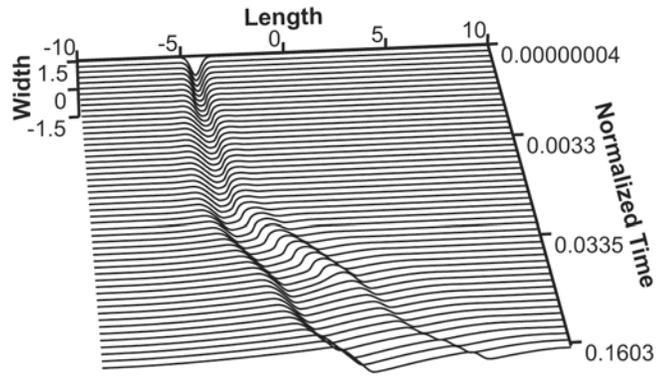

**FIG. 18.** The sidewall morphological evolution of a (111) plane exposed to the low EWI. The saw tooth shape of wave train is generated with increasing intensity in time. $V_S = 33$.

The tilt angle 45 degrees in six fold symmetry represents strong instability, where the growth rate becomes maximum. This instability is characterized by $D_\phi(0,\phi,m=3)>0$ and $D_{\phi\phi}(0,\phi,m=3)=0$, and labeled as 'D' in Fig. (2). The Gaussian edge void in this regime transform into oscillatory waves. These waves start to appear with smaller amplitudes and they spread in all directions while enhancing their strength. Meanwhile, the distorted edge-void with decreasing in strength drags towards the cathode end carrying the all package of subsidiary oscillations. This resembles the light source making translational motion while constantly emanating light waves in all directions. There is one difference, the source strength for the present case increases with time because system is regenerative. This is a good example for the behavior of a completely regenerative nonlinear system, where the source even appealingly decaying in size itself (better to say the broadening in space), it still soaks constant energy from the blowing electron winds!!!

## IV. CONCLUSIONS:

This computer simulation experiments show that the degree of symmetry of the surface, represented by the fold number in the anisotropic crystal structure, is an extremely important factor in the determination of the morphological evolution of the sidewalls as well as the life time of thin film metallic single crystal interconnect lines. In addition to this prime factor, the orientation of the principal axis of the diffusion dyadic with respect to the direction of the electric field, which is represented by the tilt angle in our formulation, also plays a significant role in the final developments in the sidewall surface topology and even in the creation of the electromigration induced internal voids, which eventually cause the fatal break down in the electrical connection either hitting the opposite sidewall or accumulating at the cathode end and resulting detrimental voiding at the contact area.

Since the present computer simulation experiments have been devoted for the electromigration dominating regime, namely $\chi > 5$, one can easily differentiate two distinct domains, (dissipative and regenerative) in terms of the tilt angle only, without going any complications with the capillary effects.**Error! Bookmark not defined.** There are two the important exceptions to this statement, which are related to the transitions taking place at the border lines while passing from one regime to another. These critical transitions in the tilt angle are corresponding to the extremum in the anisotropic diffusion coefficient denoted by $\phi = 0$, and $\phi = \pi/m$, and plus their periodic extensions. The stability and instability regimes may be defined by specifying certain bounded but open sub-intervals of the tilt angle: ($0 < \phi < \pi/2m$) and ($\pi/2m < \phi < \pi/m$), and plus their periodic extensions represented by $\{n\pi/m\}$, where $n \leq m$ is a set of positive integer numbers, respectively.

From the analysis of the computer simulation experiments, two technologically very important analytical expressions are deduced. The first expression deals with the time for the premature failure of the thin film interconnect lines, due to the EM induced wedge shape internal voids, generated by sidewall surface diffusion. The second expression is related to the incubation time for the appearance of the highly detrimental oscillatory wave trains spreading all the way along the specimen in rather catastrophically, which can be even generated by a highly localized small surface roughness under the action of the EW forces in the case of anisotropic surface diffusivity coupled with the combinations of very special surface textures and the tilt angles in fcc structures.

These catastrophic oscillatory wave train incubation time presented by Eq. (6) has an identical functional form compared to the second term in the mean time for failure (MTF) of interconnect line associated with the critical size internal void just detached from the grain boundary in bamboo structures under the action of the applied electric field. According to the recent findings of Ogurtani and Oren[2,3] in their computer simulations for the MTF[c], there is only minor difference between above cited two expressions appears to be in the pre-factors, namely; 0.28 in Eq. (6) is replaced by 0.75 in Eq. (73) of reference (2). As we mentioned in those papers, the critical void drift time dominates MTF in the case of large grain size samples compared to the grain boundary detachment time contribution, under the moderate and high current densities. Therefore, for the interconnect metallic lines having bamboo or single crystals thin films, the most important failure mechanisms are closely associated with either the electromigration induced internal wedge shape voids generated by the surface instabilities caused by the anisotropy in the surface diffusivity and/or the cathode voiding by the transport of material through the surface (sidewalls) drift-diffusion driven by the electromigration forces. In both cases the surface drift-diffusion, which takes place at interfacial layers or free surfaces, plays the predominant kinetic parameter for the determination of the life time. This parameter has to be controlled by some novel physicochemical processes such as alloying or utilizing certain liner or capping materials to reduce the atomic mobilities, increasing the interfacial Gibbs free energies, and/or decreasing the effective charge of the mobile species affected by the EM forces. Finally, the selection of the most proper micro-texture with respect to the direction of the applied current flow to reduce the adverse effects of the diffusion anisotropy is crucial; and it seems to be the {110} surface plane having about 45$^o$ tilt angle, which has almost absolute sidewall surface stability even at very high current densities, and operating temperatures.

---

[c] see; Eq. 73 in Ref.2, where there is a typographic error; $\sqrt{1.53kT\Omega_\sigma g_\sigma} \Rightarrow kT\sqrt{1.53\Omega_\sigma g_\sigma}$ .

## ACKNOWLEDGEMENTS

The authors wish to thank Professor William D. Nix of Stanford University for his very valuable suggestions and motivations at the very beginning of this extensive program at METU started in 1997. The authors also thank Professor Lucien N. Brush and Dr. Ersin Emre Oren of Washington University Seattle for offering kindly the rough draft of their paper on the linear instability analysis (LISA) before its publication.

# LIST OF TABLES

Table I. The edge void morphological evolutions on the sidewalls of the {110} surface of a fcc single crystal thin metallic film for various tilt angles, diffusion anisotropy constants, and the electron wind intensity parameters. The fold number is 2m=2.

| $\chi$ | $\theta$ | 0 | 45 | 90 | 135 |
|---|---|---|---|---|---|
|  | $A$ | \multicolumn{4}{c}{Wave evolution type} | | | |
| 5 | 5 | D | D | D | OS |
|  | 10 | V | D | S / G | OS |
| 10 | 5 | V | D | S | OS |
|  | 10 | V | D | S / G | OS |
| 25 | 5 | V | D | S | OS |
|  | 10 | V | D | S | OS |
| 50 | 5 | V | V | V | OS |
|  | 10 | V | V | V | OS |

D: decay, S: solitary wave, OS: oscillatory wave, V: internal void, F: rapture, and G: growth.

**Table II.** The edge void morphological evolutions on the sidewalls of the {100} surface of a fcc single crystal thin metallic film for various tilt angles, diffusion anisotropy coefficients and the electron wind intensity parameters. The fold number is 2m=4.

| $\chi$ | θ | 0 | 30 | 45 | 60 |
|---|---|---|---|---|---|
| | *A* | **Wave evolution type** | | | |
| 5 | 5 | K | D | S / G | OS |
| | 10 | K | D | S / G | OS |
| 10 | 5 | K | D | S / G | OS |
| | 10 | K / F | D | S / G | OS |
| 25 | 5 | K / V | D | V | OS |
| | 10 | K / F | D | V | OS |
| 50 | 5 | K | V | V | OS |
| | 10 | K / F | V | V | OS |

D-: decay; K-: kink solitary wave; OS:- oscillatory wave; V:- internal void formation, and F-: rapture.

**Table III.** The edge void morphological evolutions on the sidewalls of the {111} surface of a fcc single crystal thin metallic film for various tilt angles, diffusion anisotropy coefficients, and the electron wind intensity parameters. The fold number is 2m=6.

| $\chi$ | $\theta$ | 0 | 15 | 30 | 45 |
|---|---|---|---|---|---|
| | $A$ | **Wave evolution type** | | | |
| 5 | 5 | K / F | D / ST | K / G | OS |
| | 10 | K | D / ST | K / G | OS |
| 10 | 5 | K | D / ST | K | OS |
| | 10 | K / F | D / ST | K | OS |
| 25 | 5 | V | D / ST | V | OS |
| | 10 | F | D / ST | K | OS |
| 50 | 5 | V | D / ST | V | OS / V |
| | 10 | V | D / ST | K | OS |

K:- kink shape wave; OS:- oscillatory wave; ST:- step shape wave; V: -internal void formation; G-: growth; F-: rapture.

# REFERENCES

[1] T. O. Ogurtani, 2000. Irreversible thermokinetics theory of surfaces and interfaces with a special reference to triple junctions (unpublished).

[2] T. O. Ogurtani, E. E. Oren, Int. J. Solids Structure. 42, 3918 (2005).

[3] T. O. Ogurtani and E. E. Oren, J. Appl. Phys. 96, 7246 (2004).

[4] J. Krug and H. T. Dobbs, Phys. Rev. Lett. 73 (14), 1947 (1994).

[5] M. Schimchak and J. Krug, Phys. Rev.Lett.78 (2), 278 (1997).

[6] M. Schimschak and J. Krug, J. Appl. Phys. 87 (2) 695 (2000).

[7] M. R. Gungor and M. Maroudas, J. Appl. Phys. 85, 2233 (1999).

[8] T. O. Ogurtani and E. E. Oren, J. Appl. Phys. 90, 1564 (2001).

[9] E. E. Oren and T. O. Ogurtani, Mater. Res. Soc. Symp. Proc. 695, 209 (2002).

[10] T. O. Ogurtani and O. Akyildiz, J. Appl. Phys 97, 093520 (2005).

[11] T. O. Ogurtani and O. Akyildiz ,J. Appl. Phys (unpublished, 2005)

[12] R. Mark Bradley, Phys. Rev. E. 60, 3736 (1999).

[13] P. G. Drazin and R. S. Johnson, Solitons: an Introduction (Cambridge University Press, 1989), p. 9.

[14] T. O. Ogurtani, Solitons in Solids, Ann. Rev. Mater. Sci., 13, 67-89 (1983).

[15] R. K. Bullough and P.J. Caudrey ( Editors), Solitons (Springer-Verlag, 1980).

[16] M. Mahadevan, R. V. Bradley and J.-M. Debierre, Europhys. Lett., 45 (6), 680 (1999).

[17] A. Celik, *Electromigration Induced Hillock and Edge Void Dynamics*, M. Sc. thesis, METU, (2004).

[18] N. Marieb, E. Abratoxski and J. C. Bravman and M. Madden, and P. Flinn, J. Appl. Phys., 78 (2), 1026 (1994).

[19] T. O. Ogurtani, 2005. Unified theory of linear instability of anisotropic surfaces and interfaces under the capillary, electromigration, and elastostatic forces (unpublished).

[20] E. N. Yeremin, The Foundation of Chemica Kinetics (MIR Publ. Moscow, 1979) p.233.

[21] C. K. Hu, L. Gignac, R. Rosenberg, E. Liniger, Proceedings of the 1st International Conference on Semiconductor Technology (The Electrochemical Society, Pennington, NJ, 2001), 2001-17, P387.